\documentstyle[prb,aps,epsf,floats]{revtex}

\begin{document}
\renewcommand{\textfraction}{0.0}
\renewcommand{\floatpagefraction}{.7}
\setcounter{topnumber}{5}
\renewcommand{\topfraction}{1.0}
\setcounter{bottomnumber}{5}
\renewcommand{\bottomfraction}{1.0}
\setcounter{totalnumber}{5}
\setcounter{dbltopnumber}{2}
\renewcommand{\dbltopfraction}{0.9}
\renewcommand{\dblfloatpagefraction}{.7}

\draft

\twocolumn[\hsize\textwidth\columnwidth\hsize\csname@twocolumnfalse%
\endcsname

\title{Ground state properties of solid-on-solid models with
disordered substrates}

\author{Heiko Rieger}
\address{HLRZ c/o Forschungszentrum J\"ulich, 52425 J\"ulich, Germany}

\author{Ulrich Blasum}
\address{Zentrum f\"ur paralleles Rechnen, Universit\"at zu K\"oln,
50923 K\"oln, Germany}

\date{August 26, 1996}

\maketitle

\begin{abstract}

We study the glassy {\it super-rough} phase of a class of
solid-on-solid models with a disordered substrate in the limit of
vanishing temperature by means of {\it exact} ground states, which we
determine with a newly developed minimum cost flow algorithm.  Results
for the height-height correlation function are compared with
analytical and numerical predictions. The domain wall energy of a
boundary induced step grows logarithmically with system size, indicating
the marginal stability of the ground state, and the {\it
fractal} dimension of the step is estimated. The sensibility of the
ground state with respect to infinitesimal variations of the quenched
disorder is analyzed.

\end{abstract}
]

\newcommand{\bc}{\begin{center}}
\newcommand{\ec}{\end{center}}
\newcommand{\be}{\begin{equation}}
\newcommand{\ee}{\end{equation}}
\newcommand{\beqn}{\begin{eqnarray}}
\newcommand{\eeqn}{\end{eqnarray}}
\newcommand{\ba}{\begin{array}}
\newcommand{\ea}{\end{array}}

There has been much interest recently in the properties of crystal
surfaces upon disordered substrates \cite{toner,tsai,cule,scheidl}, in
particular the effect of the presence of pinning potentials on surface
roughening. Further motivation comes from the close relationship of
the corresponding disordered solid-on-solid model to the
two-dimensional vortex free XY model with random fields and to the
randomly pinned planar flux array
\cite{fisher,cardy,natter,toner2,hwa,batrouni,rieger,ledoussal,kierfeld}.
The latter is particularly relevant with respect to the
technologically important aspect of flux pinning in high-$T_c$
superconductors \cite{blatter}.

It has been demonstrated that these systems have a phase transition at
a critical temperature from a thermally rough phase into a {\it
  super-rough} phase at low temperatures. Whereas the existence of
this transition is established by now, the qualitative and
quantitative features of the glassy low temperature phase are still
debated.  Predictions of earlier renormalization group (RG)
calculations \cite{toner3,cardy,tsai,hwa} turned out to be
incompatible with results of extensive numerical simulations
\cite{batrouni,cule,rieger}. The subsequent discovery of the relevance
of replica symmetry breaking (RSB) effects in variational \cite{var}
and RG \cite{ledoussal,kierfeld} calculations lead to a variety of new
results, from which those obtained by the variational treatment are
again in disagreement with the most recent numerical
study \cite{marinari,lancaster}.

For us the situation seems to be the following: close to the
transition the discrimination between various predictions is
numerically hard because of its smallness on intermediate length
scales.  Far from the transition (i.e.\ deep in the glassy phase at
low temperatures) the picture should become much clearer, meaning that
the disorder dominated effects become stronger.  The clearest
evidence for the latter can be expected for strictly zero temperature,
when the roughness is exclusively produced by the substrate alone.
However, to reach this limit Monte Carlo simulations of glassy systems
like those we are interested in suffer from notorious equilibration
problems \cite{binder}. At zero temperature the properties of the
system under discussion are completely determined by its minimal energy
configuration or ground state. Since, as we show in this paper, it is
possible to calculate this state {\it exactly}, a detailed and
reliable picture of the zero temperature limit of the glassy phase
can be obtained.

Thus the aim of the present paper is twofold: first we calculate
numerically the zero temperature limit of the height-height
correlation function for finite systems and compare the result with
various analytical predictions and finite temperature simulations.
Second, motivated by the observation of manifestly glassy features
(like RSB, slow dynamics, metastable states) in the low temperature
phase, we ask how far concepts developed for finite-dimensional
spin glasses can be applied here and explore the nature of this ground
state in much greater detail. We investigate its stability with
respect to step-excitations, analyze this fractal boundary induced
domain wall itself and study the chaotic nature of the ground state by
an application of infinitesimal variations of the substrate heights.
These important issues have never been discussed in the present
context.

The solid-on-solid (SOS)-model we consider here is defined by the 
following Hamiltonian: 
\be H=\sum_{\langle ij\rangle} f(h_i-h_j)
\label{ham}
\ee
where $\langle ij\rangle$ are nearest neighbor pairs on a
$d$--dimensional lattice ($d=1,2$) and $f(x)$ is an arbitrary convex
($f''(x)\ge0$) and symmetric ($f(x)=f(-x)$) function , for instance
$f(x)=x^2$. Each height variable $h_i=d_i+n_i$ is the sum of an
integer particle number, which can also be negative, and a substrate
offset $d_i\in[0,1[$. For a flat substrate, $d_i=0$ for all sites $i$,
we have the well known SOS-model \cite{puresos}. The disordered
substrate is modeled by random offsets $d_i\in[0,1[$ \cite{tsai}. In
the present paper we are only interested in uncorrelated disorder,
meaning that all offsets are distributed independently. The method we
use is, however, applicable to any disorder distribution, in
particular to the case of correlated disorder (i.e.\ $[d_i d_{i+{\bf
    r}}]_{\rm av}=g(\bf r)$, with $g(\bf r)$ an arbitrary function),
too \cite{scheidl}. In what follows $[\ldots]_{\rm av}$ denotes the
disorder average.

The random offsets induce a frustration into the system in the same
sense as quenched randomness in the interaction strengths does in the
context of spin glass models \cite{binder}. The minimum energy surface
(being the set of particle numbers $\{n_i\}$ that minimize the energy
function $H$), is no longer flat ($h_i=const.$) but a highly
non-trivial object and the calculation of this ground state is a
complex combinatorial optimization problem. By introducing the height
difference variables $x_{ij}=n_i-n_j$ for links on the dual lattice we
can reformulate our task as a {\it minimum cost flow} problem
\be {\rm Minimize}\quad\sum_{\langle ij\rangle} c_{ij}(x_{ij})
\label{min_flow}
\ee
with integer $x_{ij}$ and convex, flow dependent (i.e.\ $x$-dependent)
cost functions $c_{ij}(x)=f(x-d_{ij})$, $d_{ij}=d_i-d_j$.  For this
problem we developed an efficient pseudo-polynomial algorithm which is
described in much detail in\cite{blasum}.  It is guaranteed to find
the {\it exact} ground state, typically in 15 minutes on a Sparc 20
workstation for a system with $N=256\times256$ sites. We used fixed
boundary conditions (b.c.) for the model (\ref{ham} becuase for technical
reasons\cite{blasum} the algorithm works only on a planar graph. We would
like to point out that this is the first extensive application of a
minimum cost flow algorithm to a problem in theoretical physics to our
knowledge \cite{maxflow}.

Much effort, numerically as well analytically, has been devoted to the
calculation of the height-height correlation function $C({\bf r},{\bf
  r'})=[(h_{\bf r}-h_{\bf r'})^2]_{\rm av}$ in the case $f(x)=x^2$
close to the
transition\cite{tsai,cule,cardy,natter,toner2,hwa,batrouni,rieger,ledoussal,kierfeld,toner3,var,marinari,lancaster}. The zero
temperature limit has not been investigated so far and the results we
present here are the first reported in the literature\cite{remark}. At
high temperatures ($T\gg T_g=2/\pi$) one expects $C({\bf r},{\bf
  r'})=T/2\pi \log |{\bf r}-{\bf r'}|$ in an infinite lattice. Because
of the natural bending of this logarithmic curve by any boundary
conditions this expression should be replaced\cite{marinari} for
finite lattices by $C({\bf r},{\bf r'})=T/2\pi P_L({\bf r},{\bf r'})$,
where for fixed b.c.\ the propagator is given by
\begin{eqnarray}
&&P_L({\bf r},{\bf r'})=\frac{2}{L^2}\sum_{n,m=1}^L\\
&&
\frac{[\sin(q_0xn)\sin(q_0ym)-\sin(q_0x'n)\sin(q_0y'm)]^2}
{2-\cos{q_0n}-\cos{q_0m}}\;,\nonumber
\end{eqnarray}
with ${\bf r}=(x,y)$, ${\bf r'}=(x',y')$ and $q_0=\pi/(L+1)$. As
mentioned above at low temperatures $T<T_g$ different
scenarios are currently under discussion in the literature: a Gaussian
scaling, with $C(r)$ linear in $\log(r)$, and a $\log^2(r)$ behavior.
In the first case a plot $C(r)$ versus $P_L(r)$ should yield a
straight line, in the second case one should be able to fit $C(r)$ to
a quadratic polynomial in $P_L(r)$ (see\cite{marinari}).

We define the site averaged correlation function:
$\overline{C}(r)=2/L^2\sum_{x=1}^{L/2}\sum_{y=1}^L [(h_{(x,y)} -
h_{(x+r,y)})^2]_{\rm av}$ and also the corresponding site averaged
lattice propagator $\overline{P}_L(r)$, which behaves for $L\to\infty$
and $r\ll L$ like $\overline{P}_L(r)\sim\log(r)/2\pi$. In fig.\ 1 we
show $\overline{C}(r)$ versus $\overline{P}_L(r)$ for a system size
$L=128$.  Obviously one does {\it not} get a straight line, which one
would obtain if $\overline{C}(r)\sim a_0 + a_1\overline{P}_L(r)$.
Thus the Gaussian scaling $\overline{C}(r)\propto\log(r)$ for
$r\to\infty$ can definitely be excluded on these length scales.

\begin{figure}[hbt]
\epsfxsize=\columnwidth\epsfbox{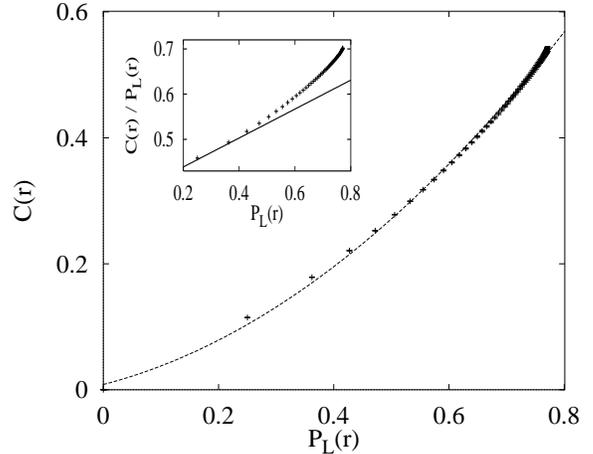}
\caption{
  The site averaged correlation function $\overline{C}(r)$ versus
  the lattice propagator $\overline{P}_L(r)$ for $L=128$ and averaged
  over 2000 samples. The broken line is a least square fit to
  $\overline{C}(r)=0.008+0.21\overline{P}_L(r)+0.57\overline{P}_L(r)^2$.
  The inset shows $\overline{C}(r)/\overline{P}_L(r)$ versus 
  $\overline{P}_L(r)$, the straight line indicates the amount of 
  curvature of the data.
}
\end{figure}

Moreover, although a fit like $\overline{C}(r)\sim a_0 +
a_1\overline{P}_L(r) + a_2\overline{P}_L(r)^2$ works fine on first sight,
there is still a significant bending in the plot of
$\overline{C}(r)/\overline{P}_L(r)$ versus $\overline{P}_L(r)$ shown in
the inset of fig.\ 1, indicating a possible presence of even higher
order terms. The fit parameters $a_1=0.21$ and $a_2=0.57$ are
compatible with a linear zero temperature extrapolation of the finite
$T$ results of Marinari et al.\ \cite{marinari}. The existence of a
$\log^2$-term is compatible with earlier \cite{cardy,toner,tsai,hwa} replica
symmetric RG calculations and with the most recent one
involving RSB \cite{ledoussal,kierfeld}. Note that the RG prediction
for the coefficient is \cite{hwa} $a_2/2\pi=2/\pi^2\cdot\tau^2+{\cal
  O}(\tau^3)$, where $\tau$ is the distance from the critical point.
Since $\tau=1$ in our case ($T=0$) it is hard to compare the numerical
value of the coefficient. Neglecting the higher order corrections
one gets $a_2/2\pi=2/\pi^2\approx0.20$ which is twice as large as
our estimate.

We also studied the correlation function for different convex cost
functions $f(x)=|x|^n$ with varying $n$. We observed that with
increasing power $n$ for smaller distances large height differences
$|x_{ij}|>1$ are suppressed due to their larger costs. At larger
distances, however, the roughness increases systematically with $n$,
expressed in monotonically increasing fit values for the coefficient
of the $\log^2$-term. This is due to the {\it decreased} costs for
small height differences ($x_{ij}=\pm1$). Moreover, we conclude that
the coefficient for the $\log^2$-term is a non-universal number at zero
temperature because of its significant dependence on the actual shape
of the cost function $f(x)$.

An intriguing question concerns the stability of the ground state with
respect to thermal fluctuations. To attack this problem we take over
the concept of domain wall renormalization that is well known in the
context of random spin systems \cite{imry,bray,riegersg}: we ask how
much energy a step (or domain wall) of height one in a system of
linear size $L$ would cost. If this energy is an increasing function
of the size $L$, it indicates the stability of the ground state even
at finite temperatures (disregarding other, more complicated
excitations). The step is induced by appropriate boundary conditions:
we fix the lower boundary to zero $(h_{(x,0)}=0$) and the upper
boundary to one $(h_{(x,L+1)}=1$).  At the left and right boundary we
enforce $h_{(0,y)}=h_{(L+1,y)}=\theta(y-L/2)$, with
$\theta(x)=0$ for $x<0$ and $\theta(x)=1$ for $x\ge0$.  This
procedure induces a straight step ($h_{(x,y)}=0$ for $y<L/2$,
$h_{(x,y)}=1$ for $y\ge L/2$) in the pure case ($d_i=0$), which costs
an energy $\Delta E=L$ (implying that the flat surface is stable at
finite temperatures in this case). In addition to quadratic geometries
we also varied the height (i.e.\ the distance between the upper and
lower boundaries) and considered rectangular geometries with $L\times
H$ sites.

We calculated the ground state first for $h_i=0$ on the whole boundary
and then for the step inducing boundaries described above (we choose
$f(x)=x^2$ from now on). In fig.\ 2 we show the averaged step energy
as a function of system size. The data are nicely fitted by a
logarithmic $L$-dependence
\be
[|\Delta E|]_{\rm av}\sim a+b\ln L\quad{\rm with}\quad b = 0.52\pm0.02 
\label{stepenergy}
\ee
At first sight the tempting conclusion of this observation would be
that there exists a {\it low temperature phase with long range order},
with an order parameter given by the overlap of the system with its
ground state (note that the boundaries are fixed so that this
definition is unambiguous). However, due to the strong chaotic
rearrangement of the state upon temperature changes, which we discuss
below, most probably this quasi long range order gets destroyed at
finite temperatures. We think that this result merits a further
investigation e.g.\ via simulations.

An inspection of the geometrical shape of the step that is induced by
the boundary condition described above reveals that it is {\it
fractal}, crossing the whole sample. Therefore we varied $H$ to ensure
that the limitation in the $y$-direction does not influence the
quantitative results we obtain.  We calculated the averaged length of
the step as a function of system size, which is depicted in the inset
of fig.\ 2, and obtain a good fit to
given by
\be 
{\cal L}_{\rm step}\propto L^{d_{\rm step}}
\quad{\rm with}\quad d_{\rm step}=1.35\pm0.02\;.
\label{fractaldim}
\ee
where $d_{\rm step}$ is the fractal dimension \cite{feder} of the step.
This is the first estimate for such an exponent in the present
context, which we expect to be universal, as we checked for various
forms for $f(x)$.

\begin{figure}[hbt]
\epsfxsize=\columnwidth\epsfbox{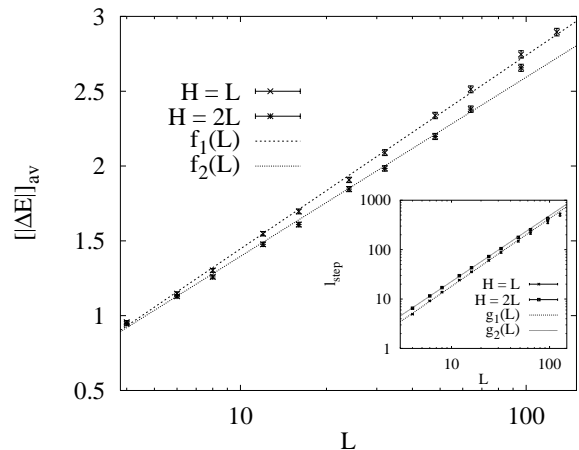}
\caption{ The averaged step energy $[|\Delta E|]_{\rm av}$ as a
  function of system size $L$, averaged over $10^4$ samples. It is
  $f_1(L)=0.15+0.56\ln(L)$ and $f_2(L)=0.19+0.52\ln(L)$, see eq.\
  (\protect{\ref{stepenergy}}).  The inset shows the step length
  ${\cal L}_{\rm step}$ as a function of system size. Here it is
  $g_1(L)=0.77\,L^{1.37}$ and $g_2(L)=1.04\,L^{1.33}$, see
  (\protect{\ref{fractaldim}}). The boxes are data for an $L\times
  2L$--geometry, the crosses for $L\times L$. Note that for larger
  $H$ the step energy is slightly smaller and the step length slighly
  larger, since the step has more space to optimize its
  configuration.}
\end{figure}

In order to shed further light on the glassy features of the
low temperature phase of our system we study what became known under
the notion of {\it chaos} in spin glasses \cite{bmchaos,rieger}.  For
the latter one observes that infinitesimal variations of temperature
or interaction strengths decorrelate the state of the system over a
distance that is called the overlap length \cite{bmchaos,rieger}. Here
we study the same scenario by infinitesimal (random) perturbations of
the offsets of the sample and comparing the resulting changes in the
ground state configuration. To be concrete we define new offsets by
$d'_i=d_i+\varepsilon_i$, with $\varepsilon_i\in[-\delta/2,+\delta/2]$
and $\delta\ll1$ and given the unperturbed and perturbed ground state
heights $h_i$ and $h'_i$, respectively, we are interested in the
accumulated height difference 
\be \chi_L(\delta)=\sum_i
[(h_i-h'_i)^2]_{\rm av}
\label{deltachi}
\ee

To get an idea about the scaling behavior of this quantity let us
consider the one-dimensional case first, for which the ground state
(with free right boundary condition) can be constructed
iteratively. The iteration $h_{i+1}=H(h_i,d_{i+1}-d_i)$ leads
asymptotically to a random walk: $[(h_{i+r}-h_i)^2]_{\rm av}\sim r$
with a slightly modified short distance behavior. The same holds for
the random variable $h_i-h'_i$ (with the site index $i$ as ``time''),
but its amplitude is reduced by factor $\delta$. Thus in one
dimension we expect $\chi_L^{(1d)}(\delta)\sim\delta^2L$, which we
verified also numerically.  Note that $\delta$ is the inverse length scale
beyond which $h_i-h'_i$ becomes typically different from zero if
$h_0=h'_0=0$.

Motivated by this observation we hypothesize the following scaling form
for $\chi_L$:
\be
\chi_L(\delta) / \delta^\eta \sim g(L)\;,
\label{2dchaos}
\ee 
with $\eta=2$ in one dimension and $g(L)$ an arbitrary function.  In
fig.\ 3 we show a scaling plot of the data we obtained from our our
ground state calculation. The data collapse is acceptable for
$\eta=0.91\pm0.05$. One should interpret $1/\delta^\eta$ as the
characteristic length scale over which the two ground states
decorrelated (in analogy to the overlap length in the context of spin
glasses \cite{bmchaos,riegersg}). We note that this is the first
quantitative description of chaos is this disordered SOS model.

\begin{figure}[hbt]
\epsfxsize=\columnwidth\epsfbox{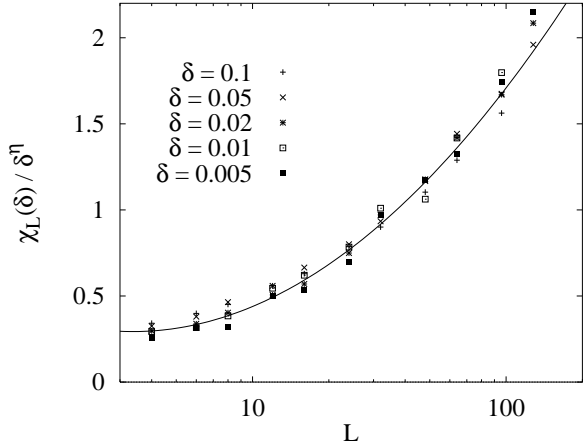}
\caption{ Scaling plot of $\chi_L^{2d}$ as defined in
  (\protect{\ref{deltachi}}) for various values for the perturbation
  amplitude $\delta$, averaged over $10^4$ samples. The best data
  collapse is obtained for $\eta=0.91$. The full line represents the
  scaling function involving a dominant $\log^2L$-term.}
\end{figure}

To summarize, our numerical investigation of the zero temperature
limit of the glassy phase in SOS models on disordered substrates
revealed the following picture: 1.) There {\it is} a dominant
$\log^2(r)$-contribution in the height-height correlation function as
predicted by RG calculations. However, we find also indications for
the existence of higher nonlinearities in $\log(r)$. 2.)  Different
forms for the microscopic interactions yield qualitatively similar but
quantitatively different results for e.g.\ the coefficient of the
$\log^2$ term (at $T=0$). 3.) the step excitation energy increases
logarithmically with the system size indicating a marginal stability of the
ground state against thermal fluctuations. 4.) The boundary induced
step itself is fractal with a fractal dimension of $d_f\approx1.35$.  
5.) The concept of the chaos known from spin glasses also applies here, the
chaos exponent for the overlap length is close to one.

As a future perspective we would like to remark that with our minimum
cost flow algorithm it is for the first time possible to attack
efficiently a large variety of combinatorial optimization problems
occuring for disordered and frustrated systems in statistical and
solid state physics, in particular those involving "real" flows like
ensembles of flux lines \cite{blatter} in lattice models
\cite{latticeflux}.

One of us (H.R.) would like to thank E.~Marinari, T.~Nattermann,
J.~J.~Ruiz--Lorenzo, G.~Batrouni, S.~Scheidl, J.~Kierfeld, P.~Le
Doussal and D.~Wolf for helpful discussions and the Deutsche
Forschungsgemeinschaft (DFG) for financial support.

\end{document}